\documentclass[superscriptaddress, amsmath, amssymb, aps, preprint]{revtex4-2}

\usepackage{graphicx}
\usepackage{dcolumn}
\usepackage{bm}
\usepackage{amsmath}
\usepackage[colorlinks,urlcolor=blue,linkcolor=blue,anchorcolor=blue,citecolor=blue]{hyperref}
\usepackage{color}

\begin{document}

\title{Directional sources realised by toroidal dipoles}

\author{Junho Jung}
\affiliation{Department of Physics, City University of Hong Kong, Tat Chee Avenue, Kowloon, Hong Kong, China}

\author{Yuqiong Cheng}
\affiliation{Department of Physics, City University of Hong Kong, Tat Chee Avenue, Kowloon, Hong Kong, China}

\author{Wanyue Xiao}
\affiliation{Department of Physics, City University of Hong Kong, Tat Chee Avenue, Kowloon, Hong Kong, China}

\author{Shubo Wang} 
\email{\textcolor{black}{shubwang@cityu.edu.hk}} 
\affiliation{Department of Physics, City University of Hong Kong, Tat Chee Avenue, Kowloon, Hong Kong, China}
\affiliation{City University of Hong Kong Shenzhen Research Institute, Shenzhen, Guangdong 518057, China \vspace{1cm}}

\begin{abstract}
\textcolor{black}{Directional optical sources can give rise to the directional excitation and propagation of light. The directionality of the conventional directional dipole (CDD) sources are attributed to the interference of the electric and/or magnetic dipoles, while the effect of the toroidal dipole on optical directionality remains unexplored.} Here, we numerically and analytically investigate the directional properties of the toroidal dipole. We show that the toroidal dipole can replace the electric dipole in the CDD sources to form the pseudo directional dipoles (PDDs), which can be applied to achieve analogous near-field directional coupling with a silicon waveguide. Moreover, the directionality of the PDDs can be flexibly controlled by changing the geometric parameters of the toroidal dipole, leading to tunable asymmetric coupling between the sources and the waveguide. These new types of directional sources provide more degrees of freedom for tailoring the optical directionality compared to the conventional sources. The results open new possibilities for directional light manipulation and can find applications in on-chip optical routing, waveguiding, and nanophotonic communications.
\end{abstract}

\maketitle

\newpage
 
\section{Introduction}
Directional scattering or propagation of light has attracted significant attention due to its application in optical sensing, switchable light routing, and optical communication 
\cite{001.PhysRevLett.100.013904,002.Liu:18,04.Jing:18, 05.10.1063/1.4934757}. Several mechanisms have been proposed to achieve the directional behaviour of light, for instance, symmetry breaking \cite{07.PhysRevB.80.035407,08,09.10.1063/1.3602322,10.Wang_Chan_2014,11.doi:10.1126/science.1233746}, nonreciprocal effect \cite{15.Shi:21,16.Shen_Zhang_Chen_Zou_Xiao_Zou_Sun_Guo_Dong_2016,17.PhysRevLett.121.153601,19.PhysRevB.100.075419,21.PhysRevB.100.245428,22.PhysRevB.105.014104,23.PhysRevB.99.245414}, multipole interference \cite{27.PhysRevLett.120.117402,29.PhysRevApplied.12.024065,30.doi:10.1073/pnas.2301620120}, etc. These mechanisms have been implemented in various optical systems for asymmetrical light manipulations. The most elemental and minimal systems are three types of directional dipoles: circular dipole, Huygens dipole, and Janus dipole, corresponding to different combinations of electric and/or magnetic dipoles \cite{27.PhysRevLett.120.117402,30.doi:10.1073/pnas.2301620120}. \textcolor{black}{These directional dipoles rely on the interference of orthogonal dipole components and exhibit different asymmetric near-field coupling with guided modes.} The circular dipole, consisting of two out-of-phase orthogonal electric or magnetic dipoles, can realise near-field directional coupling via the spin-momentum locking of the evanescent wave \cite{25.doi:10.1126/science.1233739,31.doi:10.1126/science.aaa9519}. The Huygens dipole, composed of in-phase electric and magnetic dipoles, can give rise to both the near-field and far-field directionality through the time-averaged power flow \cite{003.Geffrin_García-Cámara_Gómez-Medina_Albella_Froufe-Pérez_Eyraud_Litman_Vaillon_González_Nieto-Vesperinas_et,004.PhysRevLett.110.197401}. 
The Janus dipole, which consists of out-of-phase electric and magnetic dipoles, possesses intriguing face-dependent directional properties derived from the reactive power \cite{27.PhysRevLett.120.117402}. So far, these CDDs have been theoretically and experimentally demonstrated and widely applied in integrated optics, quantum optics, non-Hermitian optics, topological photonics, and so on \cite{33.doi:10.1126/science.1257671,34.Aiello_Banzer_Neugebauer_Leuchs_2015,37.VanMechelen:16,39.Wang_Hou_Lu_Chen_Zhang_Chan_2019,42.https://doi.org/10.1002/lpor.202000388}.

\textcolor{black}{The CDDs have limited tunability due to a lack of degrees of freedom, which hinders their application in complex optical systems.} Two main approaches have emerged to solve this problem. The first approach is to propose new directional sources for achieving flexible light routing, such as the spinning Janus dipole \cite{38.Picardi_Neugebauer_Eismann_Leuchs_Banzer_Rodríguez-Fortuño_Zayats_2019}, symmetry-engineering active meta-source \cite{02.PhysRevLett.125.157401}, and chirality-assisted directional dipole dice \cite{30.doi:10.1073/pnas.2301620120}. The second approach is to tailor the properties of the local optical fields that couple with the directional dipoles, which can be realised by tuning the structural or material properties of the systems \cite{21.PhysRevB.100.245428,42.https://doi.org/10.1002/lpor.202000388}. Here, we adopt the first approach and propose new types of directional sources consisting of the toroidal dipole, termed pseudo-directional dipoles (PDDs), which can enhance the directional light-manipulation flexibility owing to its additional internal degrees of freedom. The toroidal dipole corresponds to the lowest order of the toroidal multipole family in Cartesian multipole expansions \cite{49.PhysRevB.89.205112,43.DUBOVIK1990145,44.https://doi.org/10.1002/lpor.202200740,45.PhysRevE.65.046609}. It can be induced by the poloidal currents on a toroidal surface, equivalent to a ring of magnetic dipoles. In practice,  a toroidal dipole can be realised by using metamaterials, meta-atoms, or active current sources 
\cite{Alexey2015,51.Miroshnichenko_Evlyukhin_Yu_Bakker_Chipouline_Kuznetsov_Luk’yanchuk_Chichkov_Kivshar_2015,47.doi:10.1021/acs.nanolett.7b04200}. The toroidal dipole has been well exploited in designing optical sensors, nonradiating sources, and meta-devices \cite{006.PhysRevLett.127.096804,48.https://doi.org/10.1002/lpor.201900326,52.Yang_2019}. 
However, the role of the toroidal dipole in optical directional manipulation has not been uncovered yet, which is essential to the comprehensive understanding of toroidal physics and the discovery of new directional optical sources.  

In this article, we show that the PDDs consisting of the toroidal dipole exhibit controllable directional coupling with a dielectric waveguide. Using both full-wave finite-element simulations and analytical coupled mode theory, we show that the toroidal dipole can replace the electric dipole component(s) in conventional circular, Huygens, and Janus dipoles and give rise to similar directional near-field properties. We further investigate the tunability of the PDDs' directionality via tailoring the geometric parameters of the toroidal dipole. We demonstrate that the PDDs offer more degrees of freedom for controlling the directionality compared to the CDDs, leading to better flexibility for application in complex optical systems. 

\section{Directional coupling induced by the PDDs}

The conventional circular, Huygens, and Janus dipoles can be expressed as $\left( \pm i p \widehat{e_i}, p \widehat{e_j}\right)$, $\left( \pm m \widehat{e_i}, p \widehat{e_j}\right)$, and $\left( \pm i m \widehat{e_i}, p \widehat{e_j}\right)$, respectively, where $p$ and $m$ are the magnitudes of the electric and magnetic dipole components, respectively, with $p=m / c$, and $\widehat{e_i}$ and $\widehat{e_j}$ are the Cartesian unit axis vectors with $i \neq j \in x,y,z$ \cite{27.PhysRevLett.120.117402}. The directional properties of the CDDs originate from the interference between the dipole components, \textcolor{black}{which exhibit different symmetry properties under the parity ($P$) and time-reversal ($T$) transformations}. The electric dipole is odd under $P$ transformation and even under $T$ transformation, while the magnetic dipole is even under $P$ transformation and odd under $T$ transformation  \cite{53.PhysRevLett.113.033601}. Thus, a circular dipole is even (odd) under $P$ ($T$) transformation, a Huygens dipole is odd under both $P$ and $T$ transformations, and a Janus dipole is odd (even) under $P$ ($T$) transformation  \cite{27.PhysRevLett.120.117402}. These symmetry properties determine the directionality of the CDDs. Besides, the symmetry analysis implies that other multipoles with the same symmetry properties can give rise to analogous directional phenomena.

The toroidal dipole has different symmetry properties. It is odd under both $P$ and $T$ transformations, as shown in Table 1. Interestingly, it was shown that the toroidal dipole generates the same fields as an electric dipole does. Thus, their fields can cancel each other under the condition \cite{51.Miroshnichenko_Evlyukhin_Yu_Bakker_Chipouline_Kuznetsov_Luk’yanchuk_Chichkov_Kivshar_2015,52.Yang_2019,54.PhysRevA.92.043804}:
\begin{equation}
\mathbf{p}+i k \mathbf{t}=0,
\end{equation}
where $\mathbf{p}$ and $\mathbf{t}$ are the electric and toroidal dipole moments, respectively, and $k$ is the wavenumber. The combination of $\mathbf{p}$ and $\mathbf{t}$ satisfying the above condition corresponds to the interesting anapole state, which has zero far-field radiation. Equation (1) indicates that a toroidal dipole can replace an electric dipole to form a new set of directional dipole sources, i.e., PDDs, including the pseudo-circular dipole, pseudo-Huygens dipole, and pseudo-Janus dipole. The symmetry properties of the PDDs under $P$ and $T$ transformations are summarised in Table 1, which are the same as those of the CDDs.
\begin{table}[h!]
\caption{\label{tab:table1}%
\textcolor{black}{Symmetry properties of the dipole sources under parity ($P$) and time-reversal symmetry ($T$). The pseudo directional dipoles have the same symmetry properties as the corresponding conventional directional dipoles.}}
\begin{ruledtabular}
{\color{black}\begin{tabular}{ccc}
&$P$&$T$ \\
\hline
Electric dipole&odd&even\\
Magnetic dipole&even&odd \\
Toroidal dipole& odd & odd  \\
Pseudo-circular dipole& even & odd \\
Pseudo-Huygens dipole& odd & odd\\
Pseudo-Janus dipole& odd & even\\
\end{tabular}}
\end{ruledtabular}
\end{table}

\textcolor{black}{The three types of PDDs can be defined as:
$$
\begin{gathered}
\mathbf{D}_{\text {cir }}^{\mathrm{p}}=\left(\begin{array}{c} 
\pm i k t_{x} \\
0 \\
i k t_{z}
\end{array}\right)=\left(\begin{array}{c} 
\pm i \\
0 \\
1
\end{array}\right),
\end{gathered}
$$
$$
\begin{gathered}
\mathbf{D}_{\text {Huy }}^{\mathrm{p}}=\left(\begin{array}{c} 
0 \\
\pm m_{y}/c \\
i k t_{z}
\end{array}\right)=\left(\begin{array}{c} 
0 \\
\pm 1 \\
1
\end{array}\right),
\end{gathered}
$$
\begin{equation}
\mathbf{D}_{\text {Jan }}^{\mathrm{p}}=\left(\begin{array}{c} 
\pm m_{x}/c \\
0 \\
i k t_{z}
\end{array}\right)=\left(\begin{array}{c} 
\pm 1 \\
0 \\
i
\end{array}\right). 
\end{equation}
Here, the superscript “p” denotes “pseudo”, and the subscripts “cir”, “Huy”, and “Jan” denote “circular”, “Huygens”, and “Janus”, respectively. In the above definitions, we have assumed that the pseudo-circular dipole $\mathbf{D}_{\text {cir }}^{\mathrm{p}}$ is composed of two toroidal dipole components, corresponding to a complete pseudo-circular dipole. In addition, we can define the partial pseudo-circular dipoles of type-I $\mathbf{D}_{\text {cir }}^{\mathrm{p}}=\left(\pm ikt_{x}, 0, p_{z}\right)=\left(\pm i, 0, 1\right)$ and type-II $\mathbf{D}_{\text {cir }}^{\mathrm{p}}=\left(\pm p_{x}, 0, ikt_{z}\right)=\left(\pm i, 0, 1\right)$, which correspond to a mixing of toroidal and electric dipoles.}

To investigate the directional properties of the PDDs, we consider the dipole-waveguide system shown in Fig. 1(a). 
The silicon waveguide has a width $w =310$ nm and a height $h = 620$ nm. A PDD (enclosed by the spherical volume) is separated from the waveguide by distance $d$. The PDD is placed either on top or side of the waveguide to realise the near-field coupling, depending on the specific type of the source. We conduct finite-element simulations of the system by using COMSOL Multiphysics.  Figure 1(b) shows the dispersion relation of the guided modes of the waveguide. The waveguide supports a fundamental mode at the chosen working frequency of 108 THz. Figure 1(c) shows the eigen electric and magnetic fields of this mode at the working frequency. As seen, the electric (magnetic) field is mainly polarised in $z$ ($y$) direction.  

The toroidal dipole component in the PDDs can be realised by a ring of point magnetic dipoles, as depicted in Fig. 2(a). The ring has a radius of $r$ and consists of $n$ magnetic dipoles which are equally spaced to neighbouring dipoles. The $n$ magnetic dipoles are tangential to the ring $(\hat{\mathbf{r}} \cdot \mathbf{m}=0)$ and have equal amplitude $m$ $\left(m=\left|\mathbf{m}_1\right|=\left|\mathbf{m}_2\right|=\cdots\left|\mathbf{m}_{\mathrm{n}-1}\right|=\left|\mathbf{m}_{\mathrm{n}}\right|\right)$. In general, the toroidal dipole moment $\mathbf{t}$ resulting from Cartesian multipole expansion can be expressed as \cite{45.PhysRevE.65.046609,52.Yang_2019}
\begin{equation}
\mathbf{t}=\frac{1}{10 c} \int(\mathbf{r} \cdot \mathbf{J}) \mathbf{r}-r^2 \mathbf{J} d r,
\end{equation}
where $c$ is the speed of light in vacuum, and $\mathbf{J}$ is the poloidal current density. However, Eq. (3) cannot directly apply to our system where $\mathbf{J}$ is not explicitly defined. To circumvent this problem, we rewrite Eq. (3) in terms of magnetisation by using the relationship $\mathbf{J}=\nabla \times \mathbf{M}$:
\begin{equation}
\mathbf{t}=\frac{1}{10 c} \int[\mathbf{r} \cdot(\nabla \times \mathbf{M})] \mathbf{r}-r^2(\nabla \times \mathbf{M}) d r .
\end{equation}
For the considered system with point magnetic dipoles, the magnetisation $\mathbf{M}$ in the above equation should be replaced by $\mathbf{M} \delta\left(\mathbf{r}-\mathbf{r}_\alpha\right)$ with $ \delta\left(\mathbf{r}-\mathbf{r}_\alpha\right)$ is the three-dimensional Dirac delta function. According to the properties of the Dirac delta function, we have 
\begin{equation}
\int \mathbf{M}(r) \delta\left(\mathbf{r}-\mathbf{r}_{\alpha}\right) d r=\mathbf{m}_{\alpha}\left(\mathbf{r}_{\alpha}\right),
\end{equation}
where $\mathbf{m}_{\alpha}$ is the magnetic dipole moment located at $\mathbf{r}=\mathbf{r}_\alpha$ \textcolor{black}{and $\alpha=1,2,...,n$}. Using Eqs. (4) and (5), we can obtain the toroidal dipole moment contributed by all the point magnetic dipoles as
\begin{equation}
\mathbf{t}=\sum_{\alpha=1}^{n} \frac{\mathbf{r}_{\alpha} \times \mathbf{m}_{\alpha}}{2 c}=\frac{n r m}{2 c} \widehat{\mathbf{n}},
\end{equation}
where  $\mathbf{\widehat{n}}$ is a unit vector normal to the plane of the magnetic dipoles. The above toroidal dipole will generate the same fields as an electric dipole $\mathbf{p}$ under the condition:
\begin{equation}
m=\frac{2 c|\mathbf{p}|}{i k n r}.
\end{equation}
For simplicity, we set $|\mathbf{p}|=1$ through the paper. The toroidal dipole can be realised with different number $n$ of the discrete magnetic dipoles. To understand the effect of the values of $n$, we compute the magnetic field $|\mathbf{H}|$ inside the waveguide generated by the toroidal dipole $\mathbf{t}$.  We set $r = 0.01\lambda_{0}$ ($\approx 27.8$ nm) and $d=250$ nm. The results are shown in Fig. 2(b) as the blue symbol line for a toroidal dipole in the $x$ direction, corresponding to a ring of magnetic dipoles lying in the $yoz$ plane. The dashed black line represents the magnetic field $|\mathbf{H}|$ generated by the electric dipole $\mathbf{p}$. Figure 2(c) shows the case for a toroidal dipole in the $z$ direction, corresponding to a ring of magnetic dipoles lying in the $xoy$ plane. As seen in the two cases, when $n\le2$, the field $|\mathbf{H}|$ of the toroidal dipole $\mathbf{t}$ deviates largely from that of the electric dipole $\mathbf{p}$. For $n\ge4$, the difference between the field $|\mathbf{H}|$ of $\mathbf{t}$ and $\mathbf{p}$ becomes negligible, which implies that the toroidal dipole constructed by four magnetic dipoles can already generate the near field similar to that of the corresponding electric dipole. Thus, we will consider the toroidal dipole composed of four magnetic dipoles in the following discussions for simplicity.

 We conduct numerical simulations of the PDDs in Eq. (2) realised with the above toroidal dipole, for their near field couplings with two symmetric waveguides, as shown in Fig. 3. The PDDs are sandwiched by the two silicon waveguides. The waveguides are curved on both sides to avoid direct crosstalk between the two waveguides. The upper-right insets show the dipole arrangements. As seen, all the PDDs exhibit directional behaviour, i.e., they excite the guided waves propagating asymmetrically in the $x$, $y$ or $z$ direction. In Fig. 3(a), we notice that $\mathbf{D}_{\text {cir }}^{\mathrm{p}}$ excites the guided wave propagating to the right side of the lower waveguide and the guided wave propagating to the left side of the upper waveguide, which is similar to the conventional circular dipole. Such a phenomenon can also happen to the type-I and type-II pseudo-circular dipoles. The result for the pseudo-Huygens dipole $\mathbf{D}_{\text {Huy }}^{\mathrm{p}}$ is shown in Fig. 3(b). We see that $\mathbf{D}_{\text {Huy }}^{\mathrm{p}}$ excites the guided wave propagating to the left side of both the upper and lower waveguides, which again matches the conventional Huygens dipole. Finally, Fig. 3(c) shows the field distribution excited by the pseudo-Janus dipole $\mathbf{D}_{\text {Jan }}^{\mathrm{p}}$. The $\mathbf{D}_{\text {Jan }}^{\mathrm{p}}$ can only excite the guided waves propagating on both sides of the upper waveguide but cannot excite the guided waves in the bottom waveguide. This feature also exists in the conventional Janus dipole. In all the above cases, we observe that the strongly coupled side can be simply flipped if we change the sign of one dipole component. These results demonstrate that the PDDs can give rise to directional near-field coupling in a way similar to the CDDs.

 \section{Controlling the directionality of the PDDs}
 The directional coupling properties of the PDDs can be controlled by tailoring the geometric parameters $d$ and $r$. To understand how the two parameters affect the directional coupling quantitatively, we evaluate the directionality explicitly in both the numerical simulations and analytical theory. The numerical result of the directionality can be determined as follows:
 \textcolor{black}{
 \begin{equation}
D=\frac{\left|\Psi_{-}\right|-\left|\Psi_{+}\right|}{\left|\Psi_{-}\right|+\left|\Psi_{+}\right|},
\end{equation}
where $\Psi_{-}\left(\Psi_{+}\right)$ denotes the excited amplitude of the guided mode propagating to the negative (positive) direction of the $x$ or $z$ axis.} Analytically, we can apply the coupled mode theory to calculate the coupling coefficient of the dipole sources with the guided mode as \cite{27.PhysRevLett.120.117402,29.PhysRevApplied.12.024065}\textcolor{black}{
\begin{equation}
\kappa \propto |\mathbf{p} \cdot \mathbf{E^{*}}+\mathbf{m} \cdot \mathbf{B^{*}}|=\left|\mathbf{p} \cdot \mathbf{E^{*}}+m\sum_{\alpha=1}^{4} B^{*}_{i,j}(\mathbf{r}_{\alpha})\right|,
\end{equation}
where $\mathbf{E}$ and $\mathbf{B}$ are the electric and magnetic fields of the eigen guided mode at the dipole positions, respectively; $\sum_{\alpha=1}^{4} B^{*}_{i,j}(\mathbf{r}_{\alpha})=B^{*}_{i}(\mathbf{r}_{1})-B^{*}_{i}(\mathbf{r}_{3})-B^{*}_{j}(\mathbf{r}_{2})+B^{*}_{j}(\mathbf{r}_{4})$ is the net magnetic field acting on the toroidal dipole with $i, j \in x,y,z$ and $i \neq j$.} Then, we can analytically calculate the directionality according to the following expression
 \begin{equation}
D=\frac{\left|\kappa_{-}\right|-\left|\kappa_{+}\right|}{\left|\kappa_{-}\right|+\left|\kappa_{+}\right|},
\end{equation}
where the subscript “+” and “-” denote the propagation direction of the guided wave in the waveguide. When $|D|=1$, the coupling between the dipole sources and the waveguide is perfectly unidirectional, i.e., the guided wave propagates only in one direction. In contrast, $|D|=0$ indicates that the incident light couples to both sides of the waveguide evenly (i.e., bidirectional coupling).
 Based on Eq. (8) and Eq. (10), we now study how the geometric parameters $d$ and $r$ affect the directionality of the ideal PDDs. First, we change the separation $d$ between the dipoles and the waveguide from $d$ = 100 nm to $d$ = 500 nm while fixing the radius of the magnetic dipole ring to be $r = 0.01\lambda_{0}$ ($\approx 27.8$ nm). The numerical and analytical results of the directionality are shown in Fig. 4(a)-(c) as the green, red, and purple lines. For comparison, we also show the directionality of the CDDs as the blue lines. All the cases show good agreement between the numerical (circles) and analytical results (solid lines), demonstrating the validity of the coupled mode theory for understanding the directional coupling of the dipoles. Besides, we notice that there is no significant difference between the results of the CDDs and PDDs. This implies that the toroidal dipole and electric dipole behave similarly under the variation of the coupling distance $d$. Specifically, Fig. 4(a) shows the results for the conventional circular dipole (blue lines) and pseudo-circular dipoles, where we have considered the complete (green lines), Type-I (red lines), and Type-II (purple lines) pseudo-circular dipoles, respectively. We confirm that the three pseudo-circular dipoles induce nearly the same directionality as the conventional circular dipole, and the directionality is insensitivity to the variation of the coupling distance $d$. This is because that, in the case of the pseudo-circular dipoles, the rate of change of $\sum B^{*}_{i,j}(\mathbf{r}_{\alpha})$ and $\mathbf{E^{*}}$ with $d$ are similar, so the coupling efficiency does not change dramatically with $d$. Meanwhile, in the case of conventional circular dipole, the \textcolor{black}{$E^{*}_{x}$ and $E^{*}_{z}$} components have similar rate of change with $d$. Figures 4(b) and 4(c) show that the Huygens and Janus cases exhibit a more significant change of $D$ with varying $d$. This can be understood since the conventional Huygens and Janus dipoles depend on both the electric and magnetic fields, which have different decay rates as $d$ increases. Meanwhile, for the pseudo-Huygens and pseudo-Janus dipoles, the magnetic field \textcolor{black}{$\mathbf{B^{*}}$} and the net magnetic field $\sum B^{*}_{i,j}(\mathbf{r}_{\alpha})$ have different decay rates as $d$ increases. Therefore, the parameter $d$ can be used to tune the directionality of pseudo-Huygens and pseudo-Janus dipoles. We note that there is a trade-off between the directionality and coupling efficiency as we move the dipoles away further from the waveguide.
 
 The radius $r$ of the magnetic dipole ring can serve as another parameter to tailor the directionality of the PDDs. As a demonstration, we vary $r$ from $r = 30$ nm to $r = 230$ nm with fixed coupling distance $d = 250$ nm and magnetic dipole amplitude $m=2.39 \times 10^9$ Am$^{2}$. Figure 5 shows the directionality with respect to $r$ for the different types of PDDs, where the circles denote the numerical simulation results based on Eq. (8), while the solid lines denote the analytical results based on Eq. (10). The numerical results agree well with the analytical results. Remarkably, the directionality $D$ can be tuned for a broad range of values. All three PDDs can realise bidirectionality ($D = 0$) to near-perfect directionality ($D \approx 1$) by tailoring $r$. Specifically, Fig. 5(a) shows the directionality $D$ of the pseudo-circular dipoles (i.e., complete, type-I, and type-II pseudo-circular dipoles). We notice that different pseudo-circular dipoles exhibit different dependence on $r$. The phenomenon is attributed to both the properties of the evanescent wave and the toroidal dipole. The magnetic field of the evanescent wave is spatially nonuniform, and  the magnetic fields at the positions of the magnetic dipoles will change as $r$ increases. Additionally, the toroidal dipole moment $\mathbf{t}=\frac{nrm}{2c}\mathbf{\widehat{n}}$ will change with $r$. Together, these give rise to the different behaviours of the directionality $D$ for different pseudo-circular dipoles. Figures 5(b) and 5(c) show the directionality of the pseudo-Huygens dipole and pseudo-Janus dipole, respectively. Interestingly, the sign of $D$ can flip from positive to negative, which indicates that the strongly coupled side can be reversed without changing the direction of the dipole components. This is because the sum of magnetic fields at the positions of the magnetic dipoles (i.e., $\sum B^{*}_{i,j}(\mathbf{r}_{\alpha})$) can change sign when $r$ increases, which leads to a sign change of the \textcolor{black}{directionality $D$}. These results demonstrate the strong tunability of the near-field directionality of the PDDs.

 \section{Unidirectional coupling enabled by the optimised PDDs}
 The ideal PDDs defined in Eq. (2) cannot give rise to perfect unidirectional coupling (i.e., $|D|=1$) with the waveguide due to the polarisation mismatch. For $|D|=1$, we have $|\mathbf{p} \cdot \mathbf{E^{*}}+\mathbf{m} \cdot \mathbf{B^{*}}|=0$ for the guided mode propagating in one direction but $|\mathbf{p} \cdot \mathbf{E^{*}}+\mathbf{m} \cdot \mathbf{B^{*}}|\neq 0$ for the guided mode propagating in the opposite direction. This can be achieved by tuning the toroidal dipole moment through the magnetic dipole amplitude $m$. Since $\mathbf{E}$ and $\mathbf{B}$ fields can be numerically obtained via the eigenmode analysis with COMSOL, the value of $m$ can be calculated directly. The resulting dipoles are referred to as the optimised PDDs to differentiate them from the ideal PDDs defined in Eq. (2).
 
 Figure 6(a)-(e) shows the magnetic field $H_y$ induced by the optimised PDDs. In the simulations, we set $d$ = 250 nm and $r = 0.01\lambda_{0}$. In the cases of the optimised pseudo-circular dipoles, as shown in Figs. 6(a) and 6(b) for the type-I and type-II, respectively, the guided wave only propagates to the right side. In the case of optimised pseudo-Huygens dipole shown in Fig. 6(c), the guided wave only propagates to the left side. In the case of the optimised pseudo-Janus dipole shown in Figs. 6(d) and 6(e),  the guided waves can only be excited by one face of the dipole. These field patterns clearly demonstrate the unidirectional coupling induced by the optimised PDDs.
 
 For quantitative comparisons with the ideal PDDs, we show the field amplitude inside the waveguide generated by both the ideal and optimised PDDs, as represented by the green and blue lines in Figs. 6(f)-(i), respectively. In all the cases of the optimised PDDs, the field amplitudes at the suppressed sides of the waveguide are nearly zero, which results in a high directionality of $D\approx 0.98$. In contrast, for the ideal PDDs, although we can observe significant differences between the suppressed and strongly coupled sides, their field amplitudes are comparable to each other, especially for the ideal pseudo-Huygens and Janus cases shown in Figs. 6(h) and 6(i). We note that the optimisation of PDDs can also be achieved by tuning the radius $r$ of the toroidal dipole. However, this requires the explicit expression for the $\mathbf{r}$-dependent evanescent fields of the system to determine the optimised PDDs.

 \section{Conclusion}
 In summary, we propose new types of directional dipole sources, i.e., the PDDs and investigate their directional near-field coupling with guided waves. We demonstrate that a toroidal dipole can play the same role as an electric dipole component in the CDDs, which can give rise to similar near-field directionality. Then, we show the tunability of PDDs by varying the geometric parameters of the toroidal dipole, leading to controllable directional coupling with the guided waves that cannot be realised in conventional systems, e.g., the free switching from bidirectional coupling to unidirectional coupling without changing the amplitudes or phases of the dipole components. The directionality can also be easily flipped without changing the direction of the dipole components. Finally, we verify that these new dipole sources can be optimised to achieve perfect unidirectional coupling. The proposed PDDs can enhance the flexibility of the near-field directional manipulation and pave new ways for designing efficient optical devices and quantum sources for on-chip applications and optical communications.\\

 \section*{Acknowledgements}
 The work described in this paper was supported by the National Natural Science Foundation of China (No. 11904306 and No. 12322416) and grants from the Research Grants Council of the Hong Kong Special Administrative Region, China (No. CityU 11308223 and No. AoE/P-502/20).

\bibliography{apssamp}

\newpage

\begin{figure}[h!]\label{Fig1}
\centering
\includegraphics[width=0.45\linewidth]{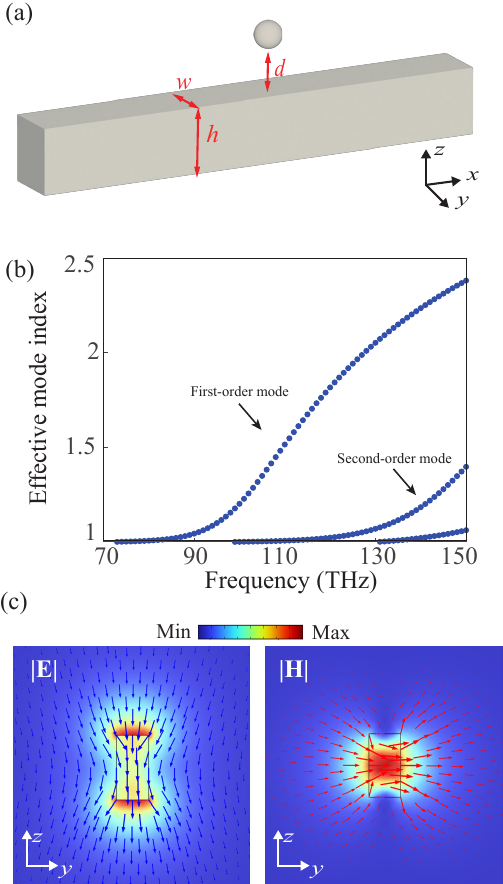} 
\caption{(a) System configuration. The dipole source is separated by $d$ from the rectangular silicon waveguide. (b) Mode dispersion relation of the silicon waveguide. (c) $|\mathbf{E}|$ (left) and $|\mathbf{H}|$ (right) field distribution and their directions (blue and red arrows) in the waveguide at the working frequency.}
\end{figure}

\begin{figure}[h!]\label{Fig2}
\centering
\includegraphics[width=0.4\linewidth]{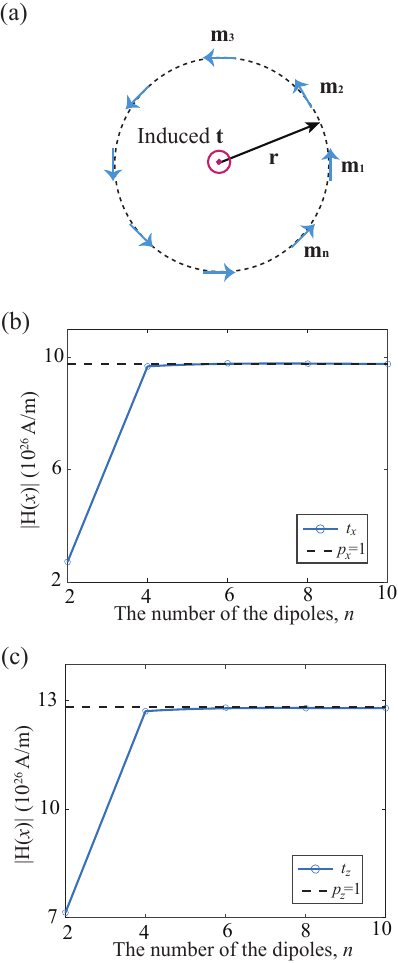} 
\caption{(a) Schematic of magnetic dipole arrangement to induce a toroidal dipole. There are $n$ number of magnetic dipoles sitting on a plane and form a ring of radius $r$, which give rise to a toroidal dipole moment perpendicular to the plane. (b-c) $|\mathbf{H}|$ field in the waveguide (at $x$ = 6000 nm) as a function of the number of magnetic dipoles forming toroidal dipole (b) $t_{x}$ and (c) $t_{z}$. The black dashed line denotes the field excited by the corresponding electric dipole with unit amplitude.}
\end{figure}

\begin{figure}[h!]\label{Fig3}
\centering
\includegraphics[width=0.5\linewidth]{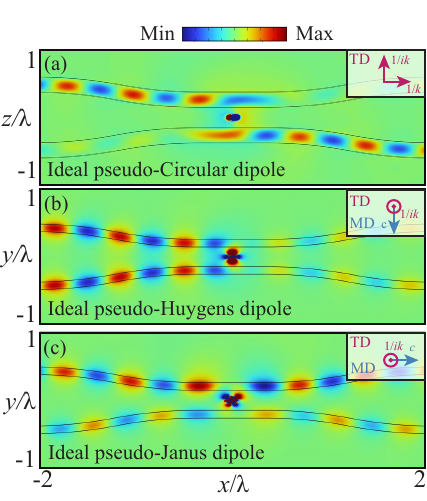} 
\caption{$H_{y}$ field distribution induced by the (a) pseudo-circular, (b) pseudo-Huygens, and (c) pseudo-Janus dipole sources. The dipole sources are sandwiched by two identical bent waveguides. The insets show dipole orientation and amplitude for each source.}
\end{figure}

\begin{figure}[h!]\label{Fig4}
\centering
\includegraphics[width=0.45\linewidth]{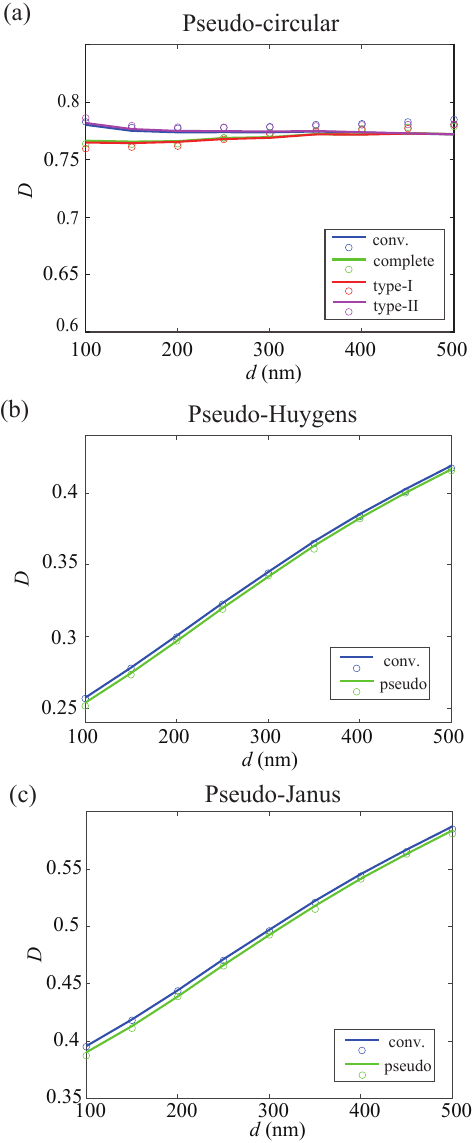} 
\caption{Directionality $D$ of the PDDs as a function of the separation $d$ between the source and waveguide. We fix $r$ = 0.01$\lambda_{0}$ ($\approx 27.8$ nm). The circles and solid lines represent the simulation and analytical results, respectively. (a) The directionality of the conventional circular dipole (blue), the complete (green), type-I (red), and type-II (purple) pseudo-circular sources. (b) The directionality of the conventional Huygens dipole (blue) and the pseudo-Huygens dipole (green). (c) The directionality of the conventional Janus dipole (blue) and the pseudo-Janus dipole (green).}
\end{figure}

\begin{figure}[h!]\label{Fig5}
\centering
\includegraphics[width=0.45\linewidth]{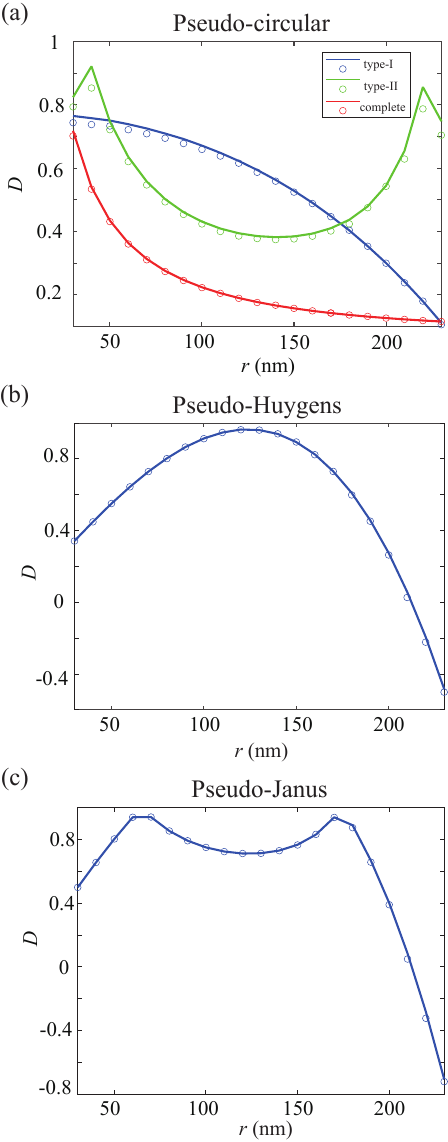} 
\caption{Directionality $D$ of the PDDs as a function of the radius $r$ of the magnetic dipole ring for (a) pseudo-circular dipoles, (b) pseudo-Huygens dipole, and (c) pseudo-Janus dipole. The circles and solid lines represent simulation and analytical results, respectively. We fix $d$ = 250 nm.}
\end{figure}

 \begin{figure*}[h!]\label{Fig6}
\centering
\includegraphics[width=\linewidth]{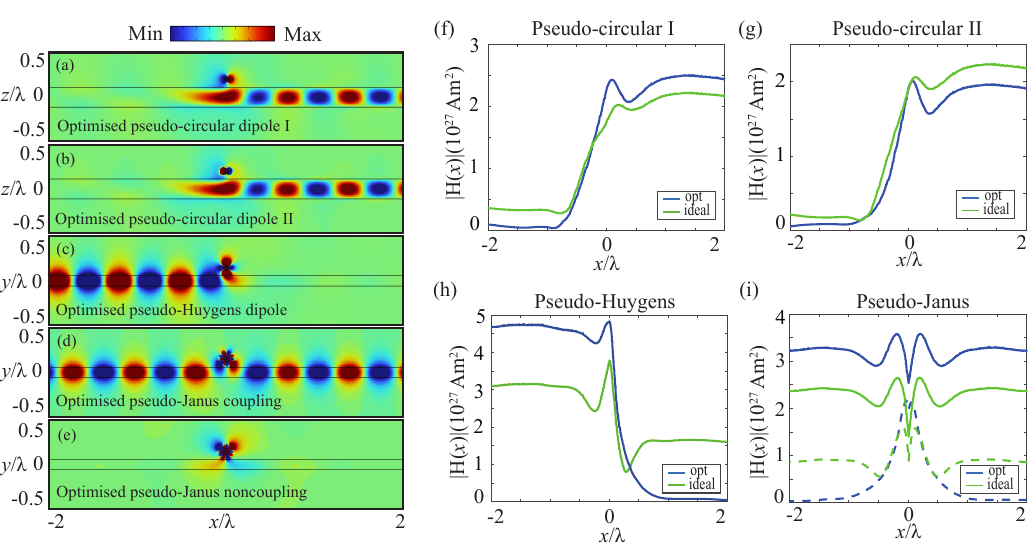} 
\caption{Unidirectional coupling induced by the optimised PDDs. The $H_{y}$ distribution generated by (a) type-I pseudo-circular source with $m=3.10\times10^{9} $Am$^{2}$, (b) type-II pseudo-circular source with $m=1.85\times10^{9}i $Am$^{2}$, (c) pseudo-Huygens source with $m=7.46\times10^{9} $Am$^{2}$, and (d,e) pseudo-Janus source with $m=5.12\times10^{9} $Am$^{2}$. We set $d$ = 250 nm and $r$ = 0.01$\lambda_{0}$. The magnetic field amplitude along the waveguide generated by (f) type-I and (g) type-II pseudo-circular dipoles. The optimised and ideal cases are denoted by blue and green solid lines, respectively. (h) Magnetic field amplitude along the waveguide generated by the optimised (blue solid line) and ideal (green solid line) pseudo-Huygens sources. (i) Magnetic field amplitude along the waveguide generated by optimised (blue) and ideal (green) pseudo-Janus sources. The solid line represents the coupling face while the dashed line represents the noncoupling face.}
\end{figure*}

\end{document}